\documentclass[11pt,twoside]{article}

\usepackage{asp2004}
\usepackage{epsf}
\usepackage{psfig}
\usepackage{lscape}

\begin{document}

\title{FUSE Spectroscopy of the DAO-type Central Star LS\,V+4621:                    
       Looking for the Photosphere in the Sea of Interstellar Absorption}
\author{M\@. Ziegler, T\@. Rauch, K\@. Werner}
\affil{Institut f\"ur Astronomie und Astrophysik, Universit\"at T\"ubingen, Sand 1, 72076 T\"ubingen, Germany}
\author{J.W\@. Kruk, C\@. Oliveira}
\affil{Department of Physics and Astronomy, Johns Hopkins University, Baltimore, MD 21218, U.S.A.}

\begin{abstract}
The far-ultraviolet spectrum of the DAO White Dwarf LS\,V+4621, 
the exciting star of the possible planetary nebula Sh\,2-216, 
is strongly contaminated by absorption features from the interstellar medium (ISM). 
For an ongoing spectral analysis, we aim to extract the pure photospheric spectrum in order to identify and model 
metal lines of species which are not detectable in the near-ultraviolet wavelength range.

We have modeled the interstellar absorption precisely and considered it for the simulation of the 
FUSE (Far Ultraviolet Spectroscopic Explorer) observation.
A state-of-the-art NLTE model-atmosphere spectrum which includes 16 elements is combined with the ISM absorption and
then compared with the FUSE spectrum.    
\end{abstract}

\section{Introduction}
\label{sect:intro}

LS\,V+4621 is one of the brightest DAO white dwarfs.
Several times, it has previously been analyzed, e.g\@.
\citet{n1999} employed NLTE model atmospheres composed out of H+He and determined an effective temperature of
$T_\mathrm{eff} = 83\,\mathrm{kK}$, a surface gravity of $\log g = 6.74\,\mathrm{(cgs)}$, and 
an abundance ratio of $n_\mathrm{He}/n_\mathrm{H} = 0.0112$ (by number).
\citet{tea2005} considered the opacity of metals (C, N, O, Si) in addition and arrived at
$T_\mathrm{eff} = 93\,\mathrm{kK}$, $\log g = 6.90\,\mathrm{(cgs)}$, 
[C]\,=\,-1.0, 
[N]\,=\,-2.0,
[O]\,=\,-0.9, and
[Si]\,=\,-0.3
([x]: log abundance / solar abundances of species x).
\citet{hea2005} used the parameters of \citet{tea2005} and investigated on the Fe abundance which, as a first result,
appeared to be supersolar.

\begin{figure}[ht]
\epsfxsize=\textwidth
\epsffile{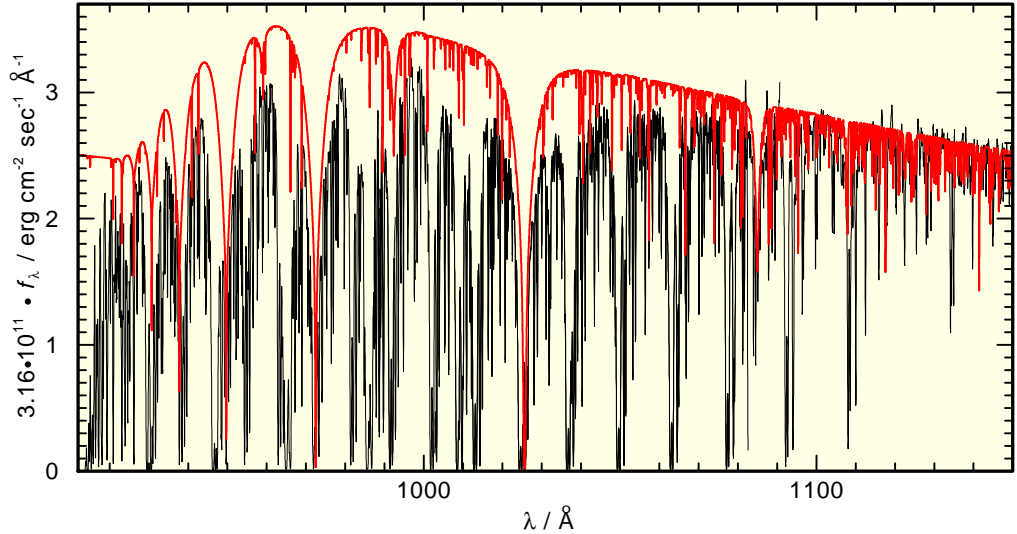}
\caption{
FUSE observation of LS\,V+4621 compared with a NLTE model-atmosphere spectrum which is normalized to match the flux level of
a HST STIS observation (cf\@. Rauch et al., these proceedings). 
The neutral hydrogen density is $\log N_\mathrm{H\,I} = 19.93\,\mathrm{(cgs)}$, and a reddening of 
$E_\mathrm{B-V} = 0.065$ is applied. Note that the deviation between observation and photospheric spectrum,
most prominent at higher energies, is due to strong ISM line absorption.
The spectra are smoothed with a Gaussian of 0.1\,\AA\ for clarity.}
\label{fig:fuse}
\end{figure}

In an on-going spectral analysis by means of NLTE model-atmosphere techniques (cf\@. Rauch et al\@. these proceedings), 
we aim to determine abundances of individual iron-group elements and to use their ionization equilibria 
in order to determine the $T_\mathrm{eff}$ precisely.
All strategic iron-group lines are located in the far and near ultraviolet and thus, we use FUSE 
and HST/STIS (Space Telescope Imaging Spectrograph) spectra for our purpose. 
Unfortunately, the FUSE wavelength region (904 - 1188\,\AA) 
is highly contaminated by interstellar absorption (Fig.~\ref{fig:fuse}), 
thus masking the photospheric information. 

\section{Spectral analysis}

In our analysis, we started to reproduce the STIS spectrum of LS\,V+4621, and the agreement
is best at $T_\mathrm{eff} = 95 \pm 2\,\mathrm{kK}$ 
and at $\log g = 6.9$.
For the abundances and a more detailed description of the analysis, 
see Rauch et al\@. (these proceedings) and \citet{zea2007}.

Since no Ca, Sc, Ti, or V lines from Kurucz's POS (i.e\@. laboratory measured) line list \citep{k1996} 
are detected in the STIS spectrum, we aim to identify these in the FUSE spectrum. 
However, this is hampered by prominent ISM absorption. 
Thus, the accurate modeling of ISM absorption is a prerequisite for a reliable analysis of the photospheric FUSE spectrum
of LS\,V+4621.

\section{ISM absorption}
\label{sect:ism}

\begin{figure}[ht]
\epsfxsize=\textwidth
\epsffile{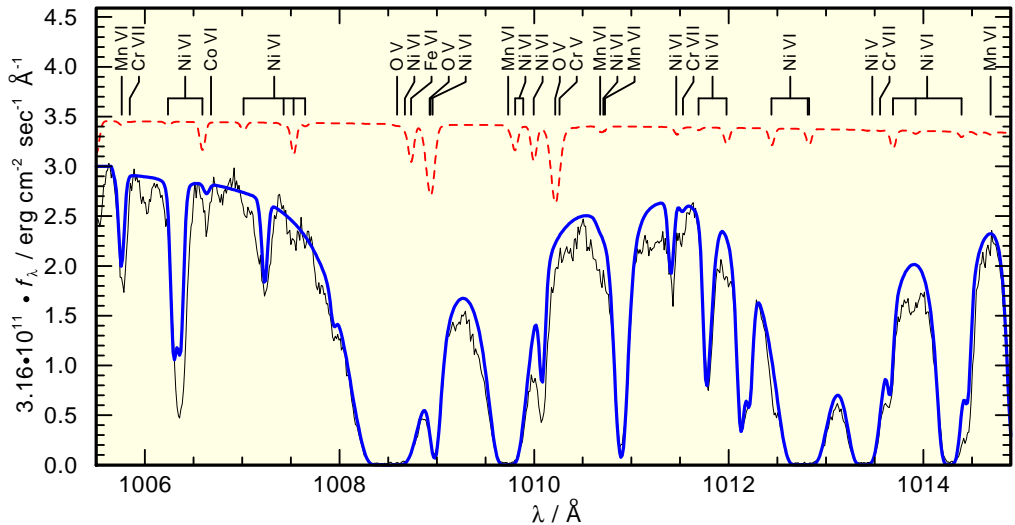}
\caption{
Section of the FUSE spectrum of LS\,V+4621 compared with our best-fitting model of the ISM absorption (thick/blue).
The thin dashed (red) line is the model-atmosphere spectrum, the marks indicate positions of photospheric lines.
Note that none of these can be identified unambiguously. However, since the pure ISM fit
is partly in very good agreement with the observation, the consideration of the model-atmosphere spectrum in the
ISM fitting process will improve the measurement of interstellar column densities.
The synthetic spectra (also in Fig.~\ref{fig:ismmod}) are convolved with a Gaussian of 0.05\,\AA\ 
in order to match FUSE's resolution.
}
\label{fig:ismnomod}
\end{figure}

We have modeled the interstellar absorption by the use of the Owens program. This takes into account
different radial and turbulent velocities, temperatures, chemical compositions, as well as column densities for each element
included \citep[see][for a detailed description]{o2007}. The comparison of our ISM model spectra with the
observation shows a good agreement (Fig.~\ref{fig:ismnomod}).

\begin{figure}[ht]
\epsfxsize=\textwidth
\epsffile{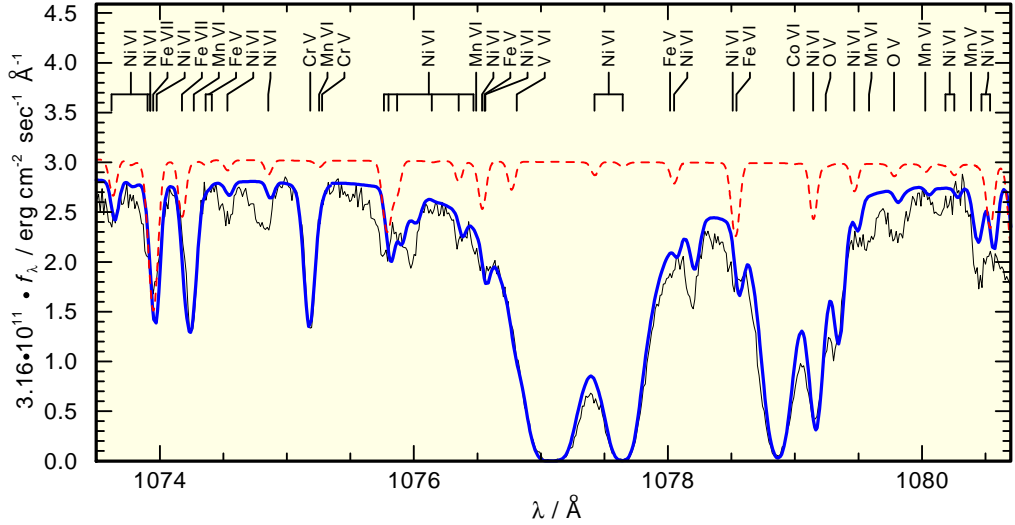}
\caption{
Section of the FUSE spectrum of LS\,V+4621 compared with the combined ISM + model-atmosphere spectrum (thick/blue).
The thin dashed (red) line is the pure model-atmosphere spectrum, the marks indicate positions of photospheric lines.
Note that is possible to identify isolated lines, e.g\@. \ion{Fe}{vii} $\lambda 1073.9\,\mathrm{\AA}$, 
which are suitable for a precise spectral analysis.
}
\label{fig:ismmod}
\end{figure}

With these results we generated a normalized ISM absorption spectrum which is then combined with the model-atmosphere spectrum. 
We are able to reproduce almost all parts of the observations and can unambiguously identify the pure photospheric
absorptions (Fig.~\ref{fig:ismmod}). These will be used to fine-tune the parameters of our model-atmosphere calculations.   

\section{Results and future work}
\label{sect:results}

We have well modeled the ISM absorption in the FUSE spectrum.
We find isolated photospheric lines which are suitable for a model-atmosphere analysis.
However, we did not succeed in our attempt to identify lines of Ca, Sc, Ti, or V so far.

The photospheric model will be included in the ISM analysis in future. 
This will enable us to calculate the impact on interstellar lines and to fine-tune and improve both, 
the ISM model as well as the photospheric model. The aim will then be to extract a pure photospheric
spectrum from the FUSE observation.

\acknowledgements T.R\@. was supported by the DLR (grants 50 OR 0201 and 05\,AC6VTB). 
M.Z\@. thanks the Astronomische Gesellschaft for a travel grant.
J.W.K\@. is supported by the FUSE project, funded by NASA contract NAS532985.
This work has made use of the line fitting procedure Owens.f developed by M\@. Lemoine and the FUSE French Team.

\end{document}